\documentclass[aps,onecolumn,showpacs]{revtex4}
\usepackage{amsmath,graphicx,amssymb,amsmath,epsfig,latexsym,wasysym,pifont,mathrsfs}
\usepackage{float} 
\usepackage{comment} 
\usepackage{verbatim}
\usepackage{lipsum}
\usepackage{hyperref}
\newcommand{\beq}{\begin{equation}}
\newcommand{\eeq}{\end{equation}}
\newcommand{\ec}{\end{center}}
\newcommand{\bc}{\begin{center}}
\newcommand{\eea}{\end{eqnarray}}
\newcommand{\bea}{\begin{eqnarray}}
\newcommand{\eeas}{\end{eqnarray*}}
\newcommand{\beas}{\begin{eqnarray*}}
\newcommand{\bd}{\begin{description}}
\newcommand{\ed}{\end{description}}
\newcommand{\bi}{\begin{itemize}}
\newcommand{\vx}{\mathbf{x}}
\newcommand{\vxp}{\mathbf{x}'}
\newcommand{\ei}{\end{itemize}}
\newcommand{\vk}{\mathbf{k}}

\newcommand{\vq}{\mathbf{q}}

\newcommand{\mysubsection}[2][]

\begin{document}
\title{Skewness in (1+1)-dimensional Kardar-Parisi-Zhang type Growth}
\author{Tapas Singha} 
\email{s.tapas@iitg.ernet.in}
\author{Malay K. Nandy}
\email{mknandy@iitg.ernet.in}
\affiliation{Department of Physics, Indian Institute of Technology
  Guwahati, Guwahati 781039,  India.}
\date{\today}

\begin{abstract}
We use the $(1+1)$-dimensional Kardar-Parisi-Zhang equation driven by
a Gaussian white noise and employ the dynamic renormalization-group of Yakhot and Orszag without rescaling [J.~Sci.\ Comput.~{\bf 1}, 3 (1986)]. 
Hence we calculate the second and third order moments of
height distribution using the diagrammatic method in the large scale and
long time limits.  The moments so calculated lead to the value $S=0.3237$ for the skewness.  This value is comparable with numerical and experimental estimates.\\\\
\end{abstract}
\maketitle
%%%%%%%%%%%%%%%%%%%%%%%%%%%%%%%%%%%%%%%%%%%%%%%%%%%%%%%%%%%%%%%%%%%%%%%%%%%%%%%%

\section{Introduction}
The study of surface growth has been one of the most important problems in 
non-equilibrium statistical physics over the past few decades
\cite{book_stanley,krug97,Halpin95,Meakin93,Family_Physica}.
The most generic continuum model of surface growth is the
Kardar-Parisi-Zhang (KPZ) equation that is endowed with interesting
properties of statistical scale invariance.  Kardar, Parisi and Zhang
\cite{KPZ86} suggested  a nonlinear differential equation for local
surface growth in the form
\beq
\frac{\partial h(\vx,t)} {\partial t}=\nu_0
\nabla^{2} h+\frac{\lambda_0}{2} (\nabla h)^{2}+ \eta(\vx,t),
\label{eq-kpz}
\eeq
where $h(\vx,t)$ is the height of the surface at position $\vx$ and time $t$ on 
a $d$ dimensional substrate, $\nu_0$ is the surface tension that has a tendency
to make the surface smooth, and the coupling constant $\lambda_0$ measures the
strength of the nonlinear interaction term. The nonlinear term induces local
growth along the normal to the surface and gives rise to lateral correlations. 
On the other hand, the linear term (containing $\nu_0$) is responsible for
diffusion of particles to the local minima \cite{VLDS_91}.  The driving term
$\eta(\vx,t)$, describing the random deposition of particles, is assumed to obey a
Gaussian distribution to account for the stochastic nature of the flux of
particles. It is taken to be a Gaussian white noise with zero mean,
$\langle\eta(\vx,t)\rangle=0$,
and with correlation
 \beq
\langle\eta(\vx,t)\,\eta(\vxp,t')\rangle=2D_{0}\,\delta^d(\vx-\vxp)\,\delta(t-t'),
\label{eq-noise}
\eeq
where $D_0$ is a constant and the angular brackets denote ensemble averages.

There are many deposition models that have been identified with the KPZ 
universality class. A few examples are the ballistic deposition
\cite{Meakin86,Family_Physica}, the Eden model
\cite{plischke_prl_84,jullien_85,eden_plischke_85}, the restricted
solid-on-solid (RSOS) model \cite{Meakin93}, and the single step model (SSM)
\cite{Meakin86,Plischek_87}.  A large number of growth experiments
show scaling exponents close to those of the KPZ growth problem. A few important
phenomena are thin film growth \cite{book_stanley}, bacterial colony
growth \cite{jullien_85,Vicsek_90}, growth of fractals \cite{Family_Vicsek91},
turbulent liquid crystal (TLC) growth \cite{K_A_Takeuchi, kazumasa}, and one
dimensional polynuclear growth (PNG) \cite{saarloos_86, goldenfeld, krug_pra_88,
Michael_Herbert_00}, etc.
Apart from such growth models, the KPZ problem is related to various other
processes such as the noisy Burgers
equation \cite{forster_pra_16_732_77}, flame front propagation 
\cite{kuramoto2003chemical,sivashinsky_Acta_77}, directed polymer in random
media \cite{KPZ86,fisher91,kpz87,JMKIM,Halpin95}, interface roughening
due to impurities \cite{huse_85,huse85}, and growing interfaces in randomly 
stirred fluids \cite{kerstein_92}.
A great amount of work has been carried out, mostly via numerical and
experimental studies, in various KPZ type surface growth problems.
The interplay of non-linearity, surface tension, and uncorrelated noise in such 
problems establish a universality class distinct from that of the
Edward-Wilkinson
type growth in the large-scale long-time limit.  The mean-square of the
height fluctuations is related to the critical exponents \cite{KPZ89} as 
\beq
\langle[h(\vx,t)- h(\vxp,t')]^{2}\rangle \sim |\vx-\vxp|^{2\chi}
\,\,\psi\left( \frac{|t-t'|}{|\vx-\vxp|^{z}} \right) , 
\eeq
where $\chi$ is the roughness exponent describing the self-affine geometry of 
the surface, $z$ is the dynamic exponent (the ratio $\frac{\chi}{z}=\beta$ is
the growth exponent), and $\psi(\cdot)$  is a scaling function.  The roughness
exponent $\chi$ is an important parameter \cite{Baiod88} in the studies of
adsorption, catalysis \cite{Pfeifer_83}, and
optical properties \cite{Moskovits_85} of a thin film. The properties of a rough
surface are determined by the distribution of
height fluctuations and it deserves attention both in
theoretical and experimental studies of growing
interfaces \cite{Meakin93}. 

Various analytical approaches have been employed to study the
universality class of the KPZ equation on the basis of scaling
exponents in different dimensions. The dynamic renormalization-group by Kardar, Parisi, and Zhang \cite{KPZ86} leads to the values of roughness exponent
$\chi=\frac{1}{2}$ and dynamic exponent $z=\frac{3}{2}$ at
one-loop order for the $(1+1)$ dimensional KPZ equation. Motivations in the theoretical study of the KPZ equation in higher dimensions have led 
to formulations of different analytical techniques.  Examples of such theoretical studies are the mode coupling scheme \cite{Colaiori_PRL_86_3946,Beijeren_PRL_54_2026, Hwa_PRA_44_R7873}, the operator product expansion \cite{Lassig_PRL_80_2366}, the self-consistent expansion \cite{Schwartz_Edwards_1992} and a nonperturbative renormalization group \cite{Canet_PRL_104_150601,Kloss_PRE2012} for the calculation of scaling exponents in the strong coupling regime.

These exponents have also been computed numerically considering different growth
rules. Apart from the numerical studies, many experiments have been carried out
to find these exponents.
Various experimental studies \cite{kazumasa,Huergo_radial,Huergo_10} have
indicated that the roughness
exponent is about $0.50$ and the growth exponent $0.33$ which have been
identified with 
the universality class of the $(1+1)$ dimensional KPZ type surface growth. 

Besides the critical exponents, the probability distribution function is
an important feature to classify the universality of a physical process
\cite{JMKIM}. 
In experiments, measurements of normalized moments is expected to be
more accurate than the measurement of scaling exponents \cite{Meakin93}.
Thus higher order moments can infer about the universality class in a better
way than the critical exponents \cite{Reis_pre_70_031603_05}. 
Consequently, higher order moments are very important in the
study of surface morphology and the universality class can be better
realized through the values of higher order moments and a parameter such as
skewness.  
    
The skewness has been computed numerically employing a variety of
deposition algorithms.  Krug et al.\ \cite{J-krug_92}, using the simulation in a
single step model for flat initial condition, obtained $|S|=0.28\pm0.04$ in the transient regime. Following the same model, they prepared stationary interfaces by taking uncorrelated spins ($\sigma_i=\pm1$) and obtained $|S|\approx0.33$. Pr{\"a}hofer and Spohn \cite{Michael_Herbert_00} took the polynuclear 
growth model and mapped it into a random permutation through the droplet
geometry thereby onto Gaussian random matrices to understand the dependence of
the initial conditions on height fluctuations. They inferred that the
droplet and flat substrates have the same scaling form but distinct
universal distributions.  They estimated the skewness for
three different shapes, namely, curved, flat, and stationary
self-similar in $(1+1)$ dimensions.  For the flat shape, they obtained
$S=0.2935$, for the curved shape, $S=0.2241$, and for the stationary 
self-similar case, $S=0.35941$. They proposed
an expression for the height distribution, namely $h(\vx,t)\simeq v_{\infty} t +
(\Gamma t)^{1/3} \zeta$ with $\zeta$ a random variable, where  $\Gamma=\frac{D_0^2\lambda_0}{8\nu_0^2}$ is a model parameter and $v_{\infty}$ is
the growth rate in the asymptotic limit \cite{K_A_Takeuchi}.  It was
found that $\zeta$ obeys the Tracy-Widom (TW) distribution
corresponding to the largest eigenvalues of random
matrices \cite{kazumasa}.  For curved interfaces the random matrices
form a Gaussian unitary ensemble (GUE) \cite{sasamoto_prl_104_230602_10}
whereas for flat interfaces they form a Gaussian orthogonal ensemble (GOE).

In an experiment on growing interfaces in liquid crystal turbulence, Takeuchi \emph{et al.} \cite{kazumasa, K_A_Takeuchi} found that the
growth and roughness exponents are the same as those of the KPZ type growth in one
dimension in the asymptotic limit. Their experimental data indicated
the value for skewness $S\simeq 0.29$ for a flat interface, whereas for
a curved interface their
experimental data converged to $S=0.2241$. They concluded that the probability
distribution function (pdf) of
interface fluctuations precisely agrees with the GOE of TW distribution for
the flat interface, whereas the curved interface fluctuations
agree with the GUE of TW distribution, up to fourth order cumulants.
Sasamoto and Spohn \cite{sasamoto_prl_104_230602_10,2010JSMTE_11_013S} solved the $(1+1)$
dimensional KPZ problem with an initial condition of curved-height
profile and showed that the pdf follows the GUE of TW distribution of 
random matrices.

It may be noted that there have been very few analytical evaluations of the 
skewness and higher order moments for the KPZ type growth problem.  The one
known to the authors is a mean field calculation yielding $S=\pm0.46$ in $(1+1)$
dimensions \cite{meanfield} with the flat initial condition $h(x,0)=0$  for the transient regime.

In this work, we are interested in the KPZ growth problem for a flat interface and
seek to calculate the skewness of height fluctuations in the stationary
state. Consequently, we apply the dynamic renormalization group scheme
without rescaling to the KPZ equation. This scheme was previously employed by 
Yakhot and Orszag \cite{yakhot_j_s_comput_1_3_86} to calculate various universal
numbers in the case of hydrodynamic turbulence. This scheme enables us to 
calculate the second and third order moments of height fluctuations in a 
straightforward manner. The ensuing result for skewness is compared with the
findings of various numerical, experimental, and theoretical studies in Table 1.
 
The paper is organized as follows.  In Section 2, the renormalizaiton-group
scheme without rescaling is applied to the KPZ problem.  Section 3 outlines
the definition of statistical moments of height fluctuations and presents
calculations of the second and third order statistical moments.  Finally,
Section 4 presents a discussion and conclusion and a comparison with other
findings.

%%%%%%%%%%%%%%%%%%%%%%%%%%%%%%%%%%%%%%%%%%%%%%%%%%%%%%%%%%%%%%%%%%%%%%%%%%%%%%%%

\section{Renormalization Scheme without Rescaling}

The nonlinear dynamics described by the KPZ equation (\ref{eq-kpz})
incorporates interaction among many degrees of freedom \cite{KPZ89}. 
The complexity of such interactions among the collective set of height
fluctuations is most easily seen when we Fourier transform the height
fluctuations $h(\vx,t)$ and the driving field $\eta(\vx,t)$.  The Fourier space
is also suitable for employing the dynamic renormalization-group
techniques \cite{hohenberg_rmp_49_435_77}. The Fourier transform of the height
fluctuations $h(\vx,t)$ is expressed as
\beq
h(\vx,t)=\int\frac{d^d k\,d\omega}{(2\pi)^{d+1}}
\, h(\vk,\omega) \, e^{i(\vk\cdot\vx-\omega t)},
\label{eq-ft}
\eeq
where $d$ is the substrate dimension.  The stochastic noise $\eta(\vx,t)$ is also Fourier transformed in a similar manner.  The Fourier amplitude of the noise
fluctuations has a zero mean, $\langle\eta(\vk,\omega)\rangle=0$, and the
noise-correlation can be expressed as
\beq
\langle\eta(\vk,\omega)\,\eta(\vk',\omega')\rangle = 2D_0
\,(2\pi)^d\delta^d(\vk+\vk')\,2\pi\delta(\omega+\omega'),
\label{eq:noise-corr-fourier}
\eeq
in the Fourier space, as a consequence of Eq.~\ref{eq-noise}.
Using Eq.\ (\ref{eq-ft}), the Fourier transform of the KPZ equation
[Eq.\ (\ref{eq-kpz})] is obtained as
\begin{equation}
(-i \omega+\nu_0 k^2)\, h(\vk,\omega)=\eta(\vk,\omega)-\frac{\lambda_0}{2} 
\int\!\!\int\frac{d^d q\,d\Omega}{(2\pi)^{d+1}}\,[\vq\cdot(\vk-\vq)]
\,h(\vq,\Omega)\,h(\vk-\vq,\omega-\Omega). 
\label{KPZFT}
\end{equation}
which is in a form particularly useful for implementing the renormalization-group scheme.

\subsection{Scale Elimination}
To implement the renormalization-group scheme,
we eliminate height fluctuations $h^>(\vq,\Omega)$ belonging to the 
shell $\Lambda_0e^{-r}\leq q \leq \Lambda_0$ in the wavevector space by substituting for
$h^>(\vq,\Omega)$ in the equation for $h^{<}(\vk,\omega)$ following from
Eq.~(\ref{KPZFT}). This process generates a perturbation series in powers of the
coupling constant $\lambda_0$. Considering terms up to second order in
$\lambda_0$ yields the equation
\begin{eqnarray}
\lefteqn{
[-i \omega+\nu_0 k^2+\Sigma(\vk,\omega)]\,h^<(\vk,\omega)
=\eta^<(\vk,\omega)\nonumber } \\
&& -\frac{\lambda_0}{2} \int\!\!\int\frac{d^d q\,d\Omega}{(2\pi)^{d+1}}
\,[\vq\cdot(\vk-\vq)]\,h^<(\vq,\Omega)\,h^<(\vk-\vq,\omega-\Omega).
\end{eqnarray}
in the range $0\leq k\leq \Lambda_0 e^{-r}$ in the wavevector space, where
$\Sigma(\vk,\omega)$
is the self energy correction represented by the amputated part of the Feynman
diagram shown in Fig.~1.
\begin{figure}[H]
\begin{center}
%\vskip -4.5in
\hskip.2in\includegraphics[scale=.85]{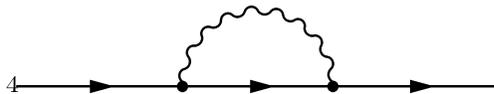}
\end{center}
%\vskip-6in
\caption{Self-energy correction. The self-energy $\Sigma(\vk,\omega)$ corresponds to the loop. Propagators are indicated by arrowed lines and correlation by a wiggly line.}
\label{fig1}
\end{figure}
The corresponding loop integral is given by
\begin{equation}
\Sigma(\vk,\omega)=  4 \left(-\frac{\lambda_0}{2}\right)^{2}
\int\frac{d^d q}{(2\pi)^d}
\, (\vk\cdot\vq) \, [\vq\cdot(\vk-\vq)]
\int^\infty_{-\infty}\frac{d\Omega}{2\pi}
\, |G^>_{0}(\hat{q})|^2 \, (2D_0) \, G^>_{0}(\hat{k}-\hat{q}),
\end{equation}
where $G_0(\hat{k}) \equiv  G_0(\vk,\omega) =[-i\omega+\nu_0 k^2]^{-1}$ is the
bare propagator and  the prefactor 4 is a combinatorial factor.
Following Refs.~\cite{KPZ89, yakhot_j_s_comput_1_3_86}, we symmetrize the
internal 
momenta by taking the transformation $\vq\rightarrow(\vq+\vk/2)$.
Performing the frequency convolution and evaluating the integral over the 
internal momenta in the shell $\Lambda_0e^{-r}\leq q\leq\Lambda_0$  yields the
self energy
\beq
\Sigma(k,0)= \frac{\lambda_0^{2} D_0}{2 \nu_0^{2}  \Lambda_0^{2-d}}
\frac{S_d}{(2\pi)^d} \left(\frac{2-d}{2d}\right)
\frac{e^{(2-d)r}-1}{2-d}\,  k^2
\eeq
in the large scale ($k\rightarrow0$) and long time ($\omega\rightarrow 0$) limits,
where $S_d=\frac{2 \pi^{d/2}}{\Gamma(d/2)}$ is the surface area of  a sphere of
unit radius embedded in a $d$ dimensional space.
As a result of the above elimination, the  effective surface
tension is obtained as
\beq
\nu^{<}(r)= \nu_0\left[1+\frac{1}{4} K_d \frac{\lambda^2_0 D_0}{\nu^3_0
\Lambda^{2-d}_0} \frac{e^{(2-d)r}-1}{d} \right],
\label{eq-nulr}
\eeq
where $K_d=\frac{S_d}{(2\pi)^d}$
and the second term in the parentheses comes from the self energy correction.

The height-height correlation is also expanded in a perturbative
series in a similar manner.  This gives rise to a correction to the noise amplitude, given by
\begin{equation} 
2D^<(r)= 2D_0+ 2 \left(\frac{-\lambda_0}{2}\right)^{2}
\int\frac{d^dq}{(2\pi)^d}
\, [\vq\cdot(\vk-\vq)]^2
\int^\infty_{-\infty}\frac{d\Omega}{2\pi}
\, |G^{>}_0(\hat{q})|^2 \, (2D_0)^2 \, |G^{>}_0(\hat{k}-\hat{q})|^2 ,
\end{equation}
where $D^{<}(r)$ is the effective amplitude of noise correlation, whereas $D_0$
is the bare parameter appearing in the
noise correlation in Eq.~(\ref{eq:noise-corr-fourier}).  The corresponding equation
is shown diagrammatically in Fig.\ 2.
\begin{figure}[H]
\begin{center}
%\vskip -4.5in
\hskip.2in\includegraphics[scale=1]{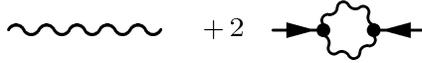}
%\vskip-6in
\end{center}
\caption{Perturbation expansion of the correlation $Q(\vk,\omega)$ to one-loop order.}
\label{fig2}
\end{figure}

Calculating the loop integral in the large scale and long time limits,
$k\rightarrow0$ and $\omega\rightarrow 0$,  the correction
to the noise amplitude is obtained as 
\beq
\Delta D = D_0\frac{\lambda_0^{2} D^2_0}{4\nu_0^{3} \Lambda^{2-d}_0}
\frac{S_d}{(2\pi)^d} \frac{e^{(2-d)r}-1}{2-d}.
\eeq 
Thus the effective amplitude of the noise correlation is given by
\beq
D^<(r)= D_0 \left [1+ \frac{1}{4} K_d \frac{\lambda_0^{2} D_0}{\nu_0^{3}
\Lambda^{2-d}_0} \frac{e^{(2-d)r}-1}{2-d} \right].
\label{eq-Dlr} 
\eeq 

We observe that the surface tension $\nu_0$ and noise amplitude $D_0$ acquire
corrections due to the elimination of small scales belonging to the
high-momentum shell
$\Lambda_0e^{-r}\leq k  \leq \Lambda_0$.

%%%%%%%%%%%%%%%%%%%%%%%%%%%%%%%%%%%%%%%%%%%%%%%%%%%%%%%%%%%%%%%%%%%%%%%%%%%%%%%%

\subsection{Flow Equations and Fixed Point}

To implement the renormalization scheme, we shall follow a procedure suggested 
by Yakhot and Orszag \cite{yakhot_j_s_comput_1_3_86,yakhot_prl_57_1772_86} where
the renormalized parameters are not rescaled after
the above scale elimination operation. A particular advantage with this scheme
is that the flow
equations for the renormalized parameters are obtained directly with
respect to the elimination parameter $r$. Implementing this scheme, we obtain,
from Eqs.~\ref{eq-nulr} and \ref{eq-Dlr}, the flow equations for the
renormalized surface tension $\nu(r)$ and renormalized noise amplitude $D(r)$ 
as the differential equations
\beq
\frac{d\nu}{dr}= \frac{1}{4} K_d \left(\frac{2-d}{d}\right) \frac{\lambda_0^{2}
D(r)}{\nu^2(r) \Lambda^{2-d}(r)} 
\label{nur}
\eeq
and
\beq 
\frac{dD}{dr}= \frac{1}{4} K_d  \frac{\lambda_0^{2} D^2(r)}{\nu^{3}(r)
\Lambda^{2-d}(r)},
\label{dr}
\eeq
where $\Lambda(r)=\Lambda_0 e^{-r}.$
In this scheme, there is no flow equation for the coupling constant
$\lambda_0$ as it does not acquire any correction due to Galilean
invariance. In order to find the fixed point, we define an effective coupling,
$g(r)$, as 
\beq 
g(r)= K_d \frac{\lambda_0^{2} D(r)}{\nu^{3}(r) \Lambda^{2-d}(r)}. 
\label{gr}
\eeq 
Using  Eqs.~\ref{nur} and \ref{dr}, the flow equation for this effective coupling is obtained as
\beq
\frac{dg}{dr}= a\,g(r)-b\,g^2(r),
\label{dgr}
\eeq
where $a=2-d$, and $b=\frac{3-2d}{2d}$.
Integrating this equation, we obtain an $r$-dependent expression for the
effective coupling, given by
\beq
g(r)=\frac{g_0 e^{a r}}{1+\frac{b}{a} g_0 (e^{ar}-1)},
\label{gr-sol}
\eeq
where $g_0=g(0)=K_d \frac{\lambda^2_0 D _0}{\nu^3_0 \Lambda^{2-d}_0}$.
The fixed point value $g^*$ is obtained in the limit $r\rightarrow\infty$.
For $d\leq 2$, we get 
\beq
 g^*=\frac{a}{b}=\frac{2d(2-d)}{(3-2d)}.
\label{fp}
\eeq
We see that the fixed point value $g^*$ diverges for the substrate
dimension $d=1.5$ and it is finite and positive in the range $0\leq d
<1.5$. However, in the range $1.5<d<2$, the coupling constant is
finite but negative, and it vanishes at $d=2$. These fixed point
values are consistent with Frey and T{\"a}uber's one-loop calculation
\cite[Cf.\ Eq.\ (3.18)]{frey_pre_50_1024_94}.
In this paper, we are interested in the substrate dimension $d=1$; thus the effective coupling constant approaches the fixed point value $g^*=2$.

Using Eqs.~(\ref{gr}) and (\ref{gr-sol}), the differential  equations (\ref{nur}) and (\ref{dr}) yield the exact solutions
\beq
\nu(r)=\nu_0\left[1+\frac{b g_0}{a} (e^{ar}-1)\right]^{a/4b d}
\eeq
and 
\beq
D(r)=D_0\left[1+\frac{b g_0}{a} (e^{ar}-1)\right]^{1/4b }.
\eeq
For very large $r$, the above solutions lead to the asymptotic expressions 
\beq 
\nu(r)\simeq \nu_0 \left(\frac{bg_0}{a}e^{ar}\right)^{a/4bd} 
\eeq
and
\beq
D(r)\simeq D_0 \left(\frac{b g_0}{a}e^{ar}\right)^{1/4b}.
\eeq
in the large scale limit.  Noting that $a=1$ and $b=\frac{1}{2}$ for our case $d=1$, these expressions for surface tension and noise amplitude reduce to
\beq
\nu(r)\simeq \nu_0 \sqrt{\frac{g_0}{2}}e^{r/2}
\label{eq-nur}
\eeq
and 
\beq
D(r)\simeq D_0 \sqrt{\frac{g_0}{2}} e^{r/2}.
\label{eq-Dr}
\eeq 
These asymptotic expressions, for very large $r$,
correspond to the renormalized surface tension
\beq
\nu(k)\simeq \nu_0 \sqrt{\frac{\lambda_0^2 D_0}{2\pi\nu^3_0}} \ k^{-1/2}
\label{eq-nu-k}
\eeq
and renormalized noise amplitude
\beq
D(k)\simeq D_0 \sqrt{\frac{\lambda_0^2 D_0}{2\pi\nu^3_0}} \ k^{-1/2}
\label{eq-D-k}
\eeq
in the large scale long time limit.

The dynamic exponent $z$ can be defined via the renormalized response function as
\beq 
G^{-1}(\mathbf{k},\omega)=\left[-i \omega+ \nu(k) k^2\right]^{-1} \propto k^{z}
\,\phi\left(\frac{\omega}{k^z}\right),
\label{eq-ren-resp}
\eeq
suggesting the scaling $\nu(k) k^2 \sim k^z$. This leads to the dynamic
exponent $z= \frac{3}{2}$ and roughness exponent $\chi=\frac{1}{2}$, the latter
being a consequence of the scaling relation $\chi+z=2$.

%%%%%%%%%%%%%%%%%%%%%%%%%%%%%%%%%%%%%%%%%%%%%%%%%%%%%%%%%%%%%%%%%%%%%%%%%%%%%%%%

\section{Statistical Moments and Skewness}

The $n$th moment of the height fluctuations is defined as
\beq
W_n=\langle[h(\vx,t)-\bar{h}(t)]^n\rangle . 
\eeq
These moments obey power laws in the stationary state and they scale as
$W_n\sim L^{n\chi}$, where $L$ is the size of the substrate.

The statistical measure corresponding to the (square of) interface width (or standard
deviation) is given by the second moment
\beq
W_2=\langle h^2(\vx,t)\rangle - \langle h(\vx,t) \rangle^2 .
\label{eq-W2}
\eeq
The skewness is related to the third moment
 \beq 
W_3=\langle
h^3(\vx,t)\rangle-3 \langle h^2(\vx,t) \rangle \langle h(\vx,t)\rangle
+ 2 \langle h(\vx,t)\rangle ^3.
\label{eq-W3}
\eeq

In this paper, we calculate the skewness $S$ of surface height fluctuations in
the KPZ surface growth model.  It is defined as
\beq
S=\frac{W_3}{(W_2)^{3/2}}.
\label{skew}
\eeq
We present the calculations of the moments $W_2$ and $W_3$ in the following subsections.

%%%%%%%%%%%%%%%%%%%%%%%%%%%%%%%%%%%%%%%%%%%%%%%%%%%%%%%%%%%%%%%%%%
\subsection{The Second Moment}

The second moment is expressed in the Fourier space as
\begin{equation}
\langle h^2(\vx,t)\rangle=
\int\frac{d^dk\,d\omega}{(2\pi)^{d+1}}
\int\frac{d^dk'\,d\omega'}{(2\pi)^{d+1}}
\,\,\langle h(\vk,\omega)\,h(\vk',\omega')\rangle
\,\,e^{i(\vk+\vk') \cdot \vx} \,\,e^{-i(\omega+\omega')t}.
\label{h^2}
\end{equation}
We shall assume the growth process to be statistically homogeneous in space and stationary in time.  This assumption yields the form
\beq
\langle h(\vk,\omega) \, h(\vk',\omega')\rangle = Q(\vk,\omega)
\,(2\pi)^d\,\delta^d(\vk+\vk')\,2\pi\,\delta(\omega+\omega')
\label{eq-hh}
\eeq
From Eq.~(\ref{KPZFT}), we see that $\langle h(\vk,\omega)\rangle =0$
for any $\vk \neq 0$, implying   $\langle h(\vx,t)\rangle =0$ for all practical
purposes. Thus from Eqs.~\ref{eq-W2}, \ref{h^2}, and \ref{eq-hh}, we obtain  
\beq
W_2=\langle h^2(\vx,t)\rangle = \int \frac{d^dk\,d\omega}{(2\pi)^{d+1}}
\,\,Q(\vk,\omega)
\label{W2FT}
\eeq
We write the integrand in terms of renormalized quantities as
\beq
W_2=\int \frac{d^{d}k\,d\omega}{(2\pi)^{d+1}}\,\, G(\vk,\omega)\, L_2(\vk,\omega)
\, G(-\vk,-\omega)
\label{eq-W2-int}
\eeq
We first consider the bare value
\begin{equation}
L_2^{(0)}(\vk,\omega)=2D_0+2\left(\frac{-\lambda_0}{2}\right)^2 \int
\frac{d^dq\, d\Omega}{(2\pi)^{d+1}}\,\,[\vq \cdot(\vk-\vq)]^2
|G_0(\vq,\Omega)|^2\,\, |G_0(\vk-\vq,\omega-\Omega)|^2 \,(2D_0)^2
\label{eq-loop2}
\end{equation}
where the propagators are unrenormalized.

We evaluate the second term in Eq.~\ref{eq-loop2} which corresponds to the amputated part of the loop diagram in Fig.~2.
Performing the integrations over the internal frequency and internal momentum in
the shell $\Lambda_0 e^{-r}\leq q \leq \Lambda_0$, we obtain 
\beq
 L^<_2(r)=2D_0 + K_d \frac{\lambda_0^2 D^2_0 }{2\nu_0^3 \Lambda_0^{2-d}}
\frac{e^{(2-d)r}-1}{2-d},
\eeq
Following Yakhot and Orszag's procedure of renormalization, we make the assumption that 
thin shells in momentum space are eliminated recursively in iterative steps. 
This leads to a differential equation for $L_2(r)$,
\beq
\frac{dL_2}{dr}=\frac{1}{2\pi}\frac{\lambda_0^2\,D^2(r)}{\nu^3(r)\,
\Lambda(r)},
\label{eq-ode-L2}
\eeq
representing the evolution of $L_2(r)$ with respect to the recursive steps of the 
shell elimination scheme.  Using Eqs.~\ref{eq-nur} and \ref{eq-Dr}, and 
integrating over $r$, Eq.~\ref{eq-ode-L2} yields
\beq
 L_2(r)=D_0\sqrt{\frac{2\lambda_0^2D_0}{\pi\nu_0^3\Lambda_0}}
\, e^{r/2}
\label{L_2}
\eeq 
for $d=1$, in the asymptotic limit of large $r$. We transform this expression
into a wavenumber and frequency dependent expression identifying
$\Lambda_0 e^{-r}$ as $ k f(\frac{\omega}{k^z})$ where $z$ is the dynamic
exponent and $f(\cdot)$ is a dimensionless scaling function. 
Thus, we obtain the renormalized function corresponding to Eq.~\ref{L_2} as 
\beq
L_2(\vk,\omega)=  D_0\sqrt{\frac{2\lambda_0^2D_0}{\pi\nu_0^3}}
%\pi (g^*)^{2} \frac{(C \nu_0)^3}{\lambda_0^2 }
\ k ^{-1/2} f^{-1/2}\left(\frac{\omega}{k^z}\right).
\eeq
We identify the scaling function by considering consistency in the $\omega
\rightarrow 0$ limit, so that
\beq
k\, f\left(\frac{\omega}{k^z}\right)= \frac{1}{ k^3\, \nu^2(k)\, |G(\vk, \omega)|^2}
\label{eq-scale-fn}
\eeq
where the modulus of the response function $G(\vk, \omega)$ signifies further consistency with the fact that $\Lambda_0 e^{-r}$ is a real quantity.
Thus the renormalized quantity $L_2(\vk,\omega)$ is expressed as
\beq
L_2(\vk,\omega)=\frac{\lambda_0^2D_0^2}{\pi\nu_0^2}
\, k \, |G(\vk,\omega)|.
\label{L2_k}
\eeq
We notice that the first diagram in Fig.~2 does not contribute to $L_2$ and the
contribution comes solely from the loop diagram.

Substituting the expression \ref{L2_k}
in Eq.~\ref{eq-W2-int} and treating the propagators as renormalized given by Eq.~\ref{eq-ren-resp}, with the renormalized surface tension $\nu(k)$ coming from Eq.~\ref{eq-nu-k}, we obtain the  contribution to the second moment as
\beq
W_2=\frac{\lambda^2_0 D^2_0}{\pi \nu^2_0} \int \frac{d^d k}{(2\pi)^d}\,\,k
\int^{+\infty}_{-\infty} \frac{d \omega}{2 \pi}
\left[\frac{1}{\omega^2+\nu^2(k)\,k^4}\right]^{3/2}
\label{eq-W2-fin}
\eeq
Performing the frequency integration using
\beq
\int^{+\infty}_{-\infty} \frac{d
\omega}{(\omega^2+m^2)^{\alpha}}=\frac{\sqrt{\pi}}{(m^2)^{\alpha-1/2}}
\frac{\Gamma(\alpha-\frac{1}{2})}{\Gamma(\alpha)}
\eeq
and carrying out the momentum integration in Eq.~\ref{eq-W2-fin}, we obtain
\beq
 W_2= \frac{4}{\pi} \left( \frac{D_0}{2\pi\nu_0} \right) \,
% (g^*)^2 \left(\frac{C \nu_0}{\lambda_0}\right)^2 \frac{1}{\pi}
\frac{1}{\mu}
\label{eq=Q2}
\eeq
where  $\mu$ is an infrared cutoff in the momentum integration.

%%%%%%%%%%%%%%%%%%%%%%%%%%%%%%%%%%%%%%%%%%%%%%%%%%%%%%%%%%%%%%%%%%%%%%%%%%%
\subsection{The Third Moment and Skewness}

The third moment $\langle h^3(\vx,t)\rangle$ can be expressed in the Fourier space as
\begin{eqnarray}
\lefteqn{
W_3=\langle h^3(\vx,t)\rangle =
\int\frac{d^dk\,d\omega}{(2\pi)^{d+1}}
\int\frac{d^dk'\,d\omega'}{(2\pi)^{d+1}}
\int\frac{d^dk''\,d\omega''}{(2\pi)^{d+1}} \nonumber}\\
&& \langle h(\vk,\omega)\,h(\vk',\omega')\,h(\vk'',\omega'')\rangle
\,\,e^{i(\vk+\vk'+\vk'') \cdot \vx}
\,e^{-i(\omega+\omega'+\omega'')t}
\label{h^3}
\end{eqnarray}
Contribution to $W_3$ comes from the one-loop diagram shown in Fig.~3.
Thus $W_3$ can be written in terms of the one-loop contribution $L_3(\hat k,\hat k')$ as 
\begin{equation}
W_3=\int \frac{d^{d+1}\hat k}{(2\pi)^{d+1}} \int
\frac{d^{d+1}\hat{k'}}{(2\pi)^{d+1}} \, G(\hat{k}) \, G(\hat{k'})
\, L_3(\hat{k};\hat{k}') \, G(-\hat{k}-\hat{k'})
\label{eq-L3-full}
\end{equation}
where $\hat k$ stands for $(\vk, \omega)$ and $\hat k'$ for
$(\vk', \omega')$.

We first consider the bare value of the loop integral.  Integrating over $\omega''$, the bare loop integral can be written as
\begin{eqnarray}
\lefteqn{
L^{(0)}_{3}(\vk,\omega;\vk',\omega')= 8 \left(\frac{-\lambda_0}{2} \right)^3 \int
\frac{d^d q\,d\Omega}{(2\pi)^{d+1}}\,\, [(\vq-\vk)\cdot(\vk'+\vk-\vq)]
\,[\vq \cdot (\vk-\vq)] \nonumber }\\
&& [-\vq \cdot(\vq-\vk'-\vk)] \, Q_0(\vq,\Omega) \,
Q_0(\vk-\vq,\omega-\Omega)\, Q_0(\vk+\vk'-\vq,\omega+\omega'-\Omega)
\label{eq-loop3}
\end{eqnarray}

\begin{figure}[H]
\begin{center}
%\vskip-0.1in
\hskip.3in \includegraphics[scale=0.5]{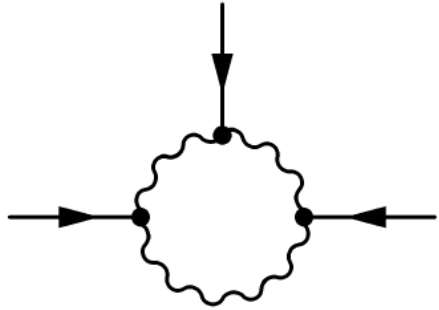}
%\vskip-0.3in
\end{center}
\caption{The third-order moment.}
\label{fig3}
\end{figure}

Carrying out the frequency convolution in Eq.~\ref{eq-loop3}, we extract the leading order contribution from this integral in the large scale and long time limits, namely the limits
$k\rightarrow 0$, $k'\rightarrow0$, $\omega\rightarrow0$, 
$\omega'\rightarrow0$ for the external momenta and frequencies. 
Working out the momentum integration  in the high-momentum shell $\Lambda_0 e^{-r} \leq q
\leq \Lambda_0$, we obtain
\beq
L^<_3(r)=\frac{3}{2} K_d \frac{\lambda^3_0 D^3_0}{\nu^5_0 \Lambda^{4-d}_0}
\frac{e^{(4-d)r}-1}{4-d}.
\eeq
As before, we consider the iterative nature of the shell elimination scheme in thin shells in the momentum space, and obtain the flow of $L_3(r)$ in the form of a differential equation
\beq
\frac{dL_3}{dr}=\frac{3}{2\pi} \frac{\lambda^3_0 D^3(r)}{\nu^5(r)}
\frac{1}{\Lambda^3(r)}
\eeq
for $d=1$.  The functions $\nu(r)$ and $D(r)$ being known from Eqs.~\ref{eq-nur} and \ref{eq-Dr}, the differential equation is solved to obtain  
\beq
L_3(r)=\frac{3}{2} \, \frac{\lambda_0 D^2_0}{\nu^2_0\Lambda_0^2} \,\, e^{2r}
\label{eq-L3r}
\eeq 
in the asymptotic limit of large $r$.

The corresponding renormalized function  $L_3(\hat k; \hat k')$, being symmetric with respect to interchange of $\hat k$ and  $\hat k'$, its frequency dependent expression can be obtained by replacing $(\Lambda_0e^{-r})^{-2}$ with the expression 
\beq
k^{-1} k'^{-1} f^{-1}\left(\frac{\omega}{k^z}\right)
f^{-1}\left(\frac{\omega'}{k'^z}\right).
\eeq
Employing Eq.~\ref{eq-scale-fn} in \ref{eq-L3r}, we thus obtain
\beq
L_3(\vk,\omega;\vk',\omega')=
\frac{3}{8\pi^2} \frac{\lambda_0^5D_0^4}{\nu_0^4}
\,\, k^{2} \, k'^{2} \, |G(\vk,\omega)|^2 \,\, |G(\vk',\omega')|^2.
\eeq
Using this expression in Eq.~\ref{eq-L3-full}, the third moment is obtained as
\begin{equation}
W_3 = \frac{3}{8\pi^2} \frac{\lambda_0^5D_0^4}{\nu_0^4}
\int\frac{d^{d+1} \hat{k}}{(2\pi)^{d+1}} \int \frac{d^{d+1}
\hat{k}'}{(2\pi)^{d+1}} \, k^2 k'^2 \, G(\hat{k}) \, |G(\hat{k})|^2
\, G(\hat{k'}) \, |G(\hat{k'})|^2 \, G(-\hat{k}-\hat{k'})
\end{equation}
where the propagators are treated as renormalized as expressed in Eq.~\ref{eq-ren-resp} with the renormalized  surface tension $\nu(k)$ given by Eq.~\ref{eq-nu-k}. Performing the frequency integrations over $\omega$ and $\omega'$  yields
\beq
W_3=\frac{3}{2}\left(\frac{D_0}{2\pi\nu_0}\right)^{3/2}
\int_{-\infty}^{+\infty} dk_x \!
\int_{-\infty}^{+\infty} dk'_x \,\, F(k_x,k_x'),
\label{eq-W3-final}
\eeq
in one dimension, where
%$A'=\frac{3}{4} \left(\frac{g^* \nu_0 C}{\lambda_0}\right)^3$ 
\beq F(k_x,k'_x)=\frac{U(k_x,k'_x)}{V(k_x,k'_x)}\label{eq-F} \eeq
with
\begin{equation}
U(k_x,k'_x)=3\left(|k_x|^3+|k'_x|^3\right)
+4|k_x+k'_x|^{3/2}\left(|k_x|^{3/2}+|k'_x|^{3/2}\right)
+ 14|k_x|^{3/2}|k'_x|^{3/2}+|k_x+k'_x|^3
\label{eq-U}
\end{equation}
and 
\beq
V(k_x,k'_x)=16\, |k_x|\,|k_x'|\,\left(|k_x|^{3/2}+|k_x'|^{3/2}+|k_x+k_x'|^{3/2}\right)^3.
\label{eq-V}
\eeq
The integrations in Eq.~\ref{eq-W3-final} can be decomposed to obtain 
\begin{equation}
W_3=\frac{3}{2}\left(\frac{D_0}{2\pi\nu_0}\right)^{3/2}
\left[ 2\int^{\infty}_{\mu} dk_x \int^{\infty}_{\mu} dk'_x \, F(k_x,k'_x)
+ 2\int^{\infty}_{\mu} dk_x \int^{\infty}_{\mu} dk'_x \, F(-k_x,k'_x)
\right]. 
\end{equation}
where we have set infrared cut offs at $\mu$ as these integrals have infrared divergences.  Thus we write
\beq
W_3=\frac{3}{2}\left(\frac{D_0}{2\pi\nu_0}\right)^{3/2}
 \left[ 2 I(\mu)+ 2 J(\mu)\right]
\label{W3}
\eeq
where 
\beq
I(\mu)= \int^{\infty}_{\mu} dk_x \int^{\infty}_{\mu} dk'_x \, F(k_x,k'_x)
\label{eq-Imu}
\eeq
and
\beq
J(\mu)=\int^{\infty}_{\mu} dk_x  \int^{\infty}_{\mu} dk'_x \, F(-k_x,k'_x).
\label{eq-Jmu}
\eeq
The infrared divergences in these integrals suggest the following forms  
\begin{eqnarray}
I(\mu)=I_0 \, \mu^{-3/2}\label{numint1}\\
J(\mu)=J_0 \, \mu^{-3/2}\label{numint2}
\end{eqnarray}
where $I_0$ and $J_0$ are dimensionless constants.

Substituting from Eqs.~(\ref{numint1}) and (\ref{numint2}) in Eq.~(\ref{W3}), we obtain the third moment as
\beq
W_3=3\left(\frac{D_0}{2\pi\nu_0}\right)^{3/2} (I_0+J_0) \, \frac{1}{\mu^{3/2}}
\label{eq-W3-fin}
\eeq
According to the definition of skewness, we thus obtain from Eqs.~\ref{eq=Q2} and \ref{eq-W3-fin}
\beq
S=\frac{W_3}{W_2^{3/2}}= \frac{3}{8} \, (I_0+J_0) \, \pi^{3/2}.
\label{eq-sk-IJ}
\eeq

We calculate the constants $I_0$ and $J_0$ from Eqs.~\ref{eq-Imu} and \ref{eq-Jmu}, using the expressions given by Eqs.~\ref{eq-F}, \ref{eq-U} and \ref{eq-V}, and obtain
\begin{eqnarray}
I_0=\lim_{\mu \rightarrow 0^+} [\mu^{3/2} I(\mu)]= 0.034946\label{numvalu1}\\
J_0=\lim_{\mu \rightarrow 0^+} [\mu^{3/2}J(\mu)]=  0.120089\label{numvalu2}
\end{eqnarray}
by means of numerical integrations.  The computation shows convergence to the above numerical values as the parameter $\mu$ is chosen to approach values close to zero.

From Eq.~\ref{eq-sk-IJ}, the value of skewness is thus found to be 
\beq
S=\frac{3}{8} \, (0.155035) \, \pi^{3/2}=0.323732. 
\label{eq-Skew-value}
\eeq

%%%%%%%%%%%%%%%%%%%%%%%%%%%%%%%%%%%%%%%%%%%%%%%%%%%%%%%%%%%%%%%%%%%%%%%%%%%%%%%%

\section{Discussion and Conclusion}

We employed Yakhot and Orszag's scheme of renormalization without rescaling and obtained the renormalized surface tension and the strength of the noise correlation for the surface growth problem governed by the KPZ dynamics on a flat substrate. This scheme of renormalization is slightly different from the usual perturbative renormalization group analysis with rescaling that has been employed for dynamical problems by Ma and Mazenko \cite{s.k.Ma_prb_11_4077_75}, Forster \emph{et al.} \cite{forster_pra_16_732_77}, and Medina \emph{et al.} \cite{KPZ89}. This method allowed us the advantage of obtaining the flow equations directly without rescaling by considering the iterative nature of the scale elimination procedure. This yielded the fixed point from the $r$-dependent expression of effective coupling constant $g(r)$ in the limit $r\rightarrow\infty$.  Similar to the other calculations, the renormalized surface tension and the strength of the noise correlation are found to be renormalized in the same way so that $D(r)/\nu(r)$ is $r$-independent, a consequence of fluctuation dissipation theorem for the case of $(1+1)$ dimensional KPZ equation \cite{forster_pra_16_732_77,Deker}.

To obtain a numerical value for the skewness,  we employed the diagrammatic approach for the second and third order moments $W_2$ and $W_3$. The Fourier integrals of these moments involve the loop integrals $L_2$ and $L_3$.  The simplicity of Yakhot and Orszag's renormalization scheme allowed us to find renormalized expressions for these loop integrals in a straightforward manner. Although the renormalized diagrams are infrared divergent, the calculated value of skewness turns out to be finite due to cancellation of the infrared cutoff parameter $\mu$.  We obtained a value of skewness $S=0.323732$ for the flat geometry in the stationary state which is compared with the results of numerical simulations for various growth models and those of experiments in Table~1. We present a discussion with regard to these results in the following paragraphs.

It has been shown by numerical simulations for polynuclear growth (PNG)
that the roughness and growth exponents are in good agreement with the
one dimensional KPZ exponents\ \cite{saarloos_86,goldenfeld}. Further
numerical work by Krug \emph{et al.}\ \cite{J-krug_92} and  Bartelt and
Evance\ \cite{bartelt_jpa_26_2743_93} have ensured that the PNG model belongs to
the universality class of the KPZ growth model \cite{Meakin93}.  Pr{\"a}hofer and
Spohn\ \cite{Michael_Herbert_00} have shown that the PNG model follows the
TW distribution with different initial conditions. They
estimated the skewness for three different shapes,
namely,  $S=0.2241$ for the curved shape (GUE TW), 
$S=0.2935$ for the flat shape (GOE TW), and $S=0.35941$ for the stationary
self-similar case \cite{Michael_Herbert_00}.  On the 
other hand, the distribution of height fluctuations for the KPZ growth model with sharp wedge initial condition was shown to be the same as that of the GUE TW distribution, as established by Sasamoto and Spohn \cite{sasamoto_prl_104_230602_10,2010JSMTE_11_013S}.
Calabrese and Doussal \cite{calabrese_prl_106_250603_11}
obtained the GOE TW distribution by mapping the one dimensional KPZ problem with flat initial condition to a one end
free directed polymer, referred to as a point-to-line configuration. The curved initial condition, on the other hand, maps on to a point-to-point configuration of the directed polymer.

For the case of directed polymers at zero temperature in a random potential
(DPRP), Kim et al.\ \cite{JMKIM}
introduced two types of random site potentials $\mu(\vx,t)$, namely, 
uniform and Gaussian distributions for $\mu(\vx,t)$, with the bending
energy ($\gamma$) of the polymer as the only tunable parameter. They
obtained skewness $S=-0.29\pm0.02$ of the minimum energy distribution in $1+1$ dimensions for uniform distribution of $\mu(\vx,t)$ for a point-to-line configuration via simulations for
$\gamma>1$. The same value of skewness was obtained for Gaussian
distribution of $\mu(\vx,t)$, which is independent of $\gamma$.
Kim and others \cite{jmkim_prl_62_2289_89,JMKIM} studied height fluctuations
of surface growth using the RSOS model with a flat initial condition where the scaling form of the height 
distribution matches with the energy
distribution of the DPRP within numerical accuracy. The skewness
in the same model turned out to be $S\approx -0.29$, suggesting universality of the probability distribution function.

Using a mean field theory in terms of densities at different heights
applied to the KPZ equation in $(1+1)$ dimensions, Ginelli and
Hinrichsen\ \cite{meanfield} started with the flat initial condition $h(x,0)=0$ and obtained the skewness $S=\pm0.46$ for the transient regime.
Takeuchi \emph{et al.}\ \cite{kazumasa} carried out an experiment on a
growing interface in liquid crystal turbulence and established
that it is in the KPZ universality  class.  For flat initial conditions, their
experimental asymptotic value for skewness was close to $\simeq 0.29$
as suggested by their experimental plots.  These values are displayed 
in Table~1 for comparison with our result.

%%%%%%%%%%%%%%%%%%%%%%%%%%%%%%%%%%%%%%%%%%%%%%%%%%%%%
 \begin{table}[ht]
\caption{Values of Skewness in one dimension}
 \begin{center}
 \begin{tabular}{l|c|l|c}
 \hline\hline
 \it System of Study & \it Reference & \it Methodology & \it Skewness \\ [.5ex] \hline
 SSM (flat) & \cite{J-krug_92} & Numerical  & $0.28\pm0.04$  \\ \hline
 SSM (stationary) & \cite{J-krug_92} & Numerical  & $\approx0.33$  \\ \hline
 DPRM (point-to-line) & \cite{J-krug_92}& Numerical   & $-0.296\pm0.028$ \\ \hline
 DPRP (point-to-line) & \cite{JMKIM}  & Numerical   & $-0.29\pm0.02$  \\ \hline
 RSOS (flat) & \cite{jmkim_prl_62_2289_89,JMKIM} & Numerical & $\approx -0.29 $\\ \hline
 TLC (flat) & \cite{kazumasa} & Experimental & $0.29$  \\ \hline
 PNG (curved) & \cite{Michael_Herbert_00}& Numerical & 0.2241  \\ \hline
 PNG (flat) & \cite{Michael_Herbert_00}& Numerical  & $0.2935$  \\ \hline
 PNG  (stationary) & \cite{Michael_Herbert_00}& Numerical  & $0.35941$  \\ \hline
 KPZ (mean field, flat) & \cite{meanfield} & Analytical  & $\pm 0.46$ \\ \hline
 %DPRM & \cite{J-krug_92} & Numerical   & $\approx -0.33$  \\ \hline
 Combustion front (flat) & \cite{miettinena_epjb_46_55_05}& Experimental & $0.33$ \\ \hline
 Combustion front (stationary) & \cite{miettinena_epjb_46_55_05}& Experimental & $0.32$ \\ \hline
 KPZ (present calculation) & Eq.~(\ref{eq-Skew-value}) & Analytical & $0.3237$ \\ [1ex] \hline 
\end{tabular}
\end{center}
\label{table:nonlin}
\end{table}
%%%%%%%%%%%%%%%%%%%%%%%%%%%%%%%%%%%%%%%%%%%%%%%%%%%%%%%%

We observe that our calculated result for the skewness is comparable with some of the experimental values and numerical simulations.  It deviates from the non-stationary results and the deviation is more pronounced from the result for curved interfaces. This is expected as our calculations are applicable for a flat geometry in the stationary state.  We also observe that the result of the mean field calculation deviates somewhat strongly from all other results.  Since skewness is determined by the underlying probability distribution, its calculation following from the governing dynamics is of importance in inferring the universality class.  Moreover, the existing studies indicate that the pdf is determined by not only the governing dynamics but also by the boundary conditions.  Thus it may be said that there are different subclasses belonging to the same universality class. Different numerical values of skewness may thus be said to correspond to different universality subclasses although they may have the same scaling exponents for the correlation and response functions.

Since the renormalization scheme involves calculations of the statistical
moments in the large scale limit $k\rightarrow0$, such calculations are
expected to lead to the statistical properties of the growth process at large scales.  The fact that $W_2$ and $W_3$
turn out to be infrared divergent implies a dominant role of the large scale
fluctuations in determining these statistical moments.   Moreover since the renormalization scheme involves calculations in the long time limit $\omega\rightarrow0$, such calculations are expected to capture the statistical properties of the growth process in the stationary state. 

However, for a large system, achieving a stationary state is
difficult \cite{Imamura_prl_108_2012}, especially in experiments and numerical
simulations, unless a stationary state is taken as an initial condition
\cite{Takeuchi_prl_110_2013}.  To achieve a stationary pdf for the flat one dimensional KPZ problem, Imamura and Sasamoto took both sided Brownian motion as an initial condition \cite{Imamura_prl_108_2012,Imamura_JSP_2013} and obtained the generating function for the replica partition function as a Fredholm determinant.  This allowed for the calculation of the pdf which was found to approach the Baik-Rains $F_0$ distribution in the long time limit.
Krug \emph{et al.} \cite{J-krug_92}
investigated the stationary state skewness for the SSM model with random uncorrelated spins ($\sigma_i=\pm1$).  They obtained $|S|\approx 0.33$ which agrees well with our calculated value.
Maunuksela \emph{et al.} \cite{Maunuksela_PRL_79_1515} identified that the
universality class of slow combustion fronts of a paper sheet belongs to the
KPZ universality class on the basis of scaling exponents.
With the same experimental conditions, Miettinena \emph{et al.\/}
\cite{miettinena_epjb_46_55_05} performed an experiment on paper burning
to find the skewness from the pdf. They studied the height distribution of combustion fronts
for flat initial conditions in the saturation regime and obtained the value
$S=0.32$ which agrees well with our calculated value. It may however be noted
that Takeuchi \cite{coment_Takeuchi_12} has suggested that their analysis of
the pdf may need modifications.  

Although our calculated value for the skewness ($0.3237$) compares excellently well (within 1--2\%) with the above mentioned stationary values ($0.33$ and $0.32$), we observe that there is a slight departure (of about $10\%$) from the value $S=0.3594$ coming from the PNG model \cite{Michael_Herbert_00}.  This departure may be due to the dominating role of large scale fluctuations in determining the moments $W_2$ and $W_3$.  The infrared cutoff $\mu$ may be interpreted as the inverse of the size $L$ of the substrate and thus the calculations appear to be influenced by finite size effects.  In spite of this slight departure, together with the agreements with the experimental results, our calculation for the skewness seems to identify the relevant universality subclass of the KPZ equation.

Ideally speaking, full information about the pdf enables one to classify many
seemingly similar problems into varying universality classes. However, an
analytical calculation of the full pdf is an extremely difficult task, whereas the
calculation of higher order moments such as the skewness is a more viable
approach.  Thus a classification scheme for universality beyond the
scaling exponents can be formulated via the values of skewness for various
processes.  Takeuchi and Sano \cite{K_A_Takeuchi} have proved through the TLC experiment that the KPZ class has geometry dependent subclasses in spite of having the same scaling exponents. Thus, to characterize the subclasses, knowledge of skewness and higher order moments is very much essential. The calculation of skewness, directly from the KPZ dynamics, is therefore an important step in identifying a universality subclass of the KPZ equation.

%%%%%%%%%%%%%%%%%%%%%%%%%%%%%%%%%%%%%%%%%%%%%%%%%%%%%%%%%%%%%%%%%%%%%%%%%%%%


\begin{thebibliography}{10}

\bibitem{book_stanley}
A.-L. Barab{\'a}si and H.~E. Stanley,
\newblock {\em Fractal Concepts in Surface Growth},
\newblock (Cambridge University Press, Cambridge, England, 1995).

\bibitem{krug97}
J.~Krug,
\newblock Adv. Phys. {\bf 46}, 139 (1997).

\bibitem{Halpin95}
T.~Halpin-Healy and Y.-C. Zhang,
\newblock Phys. Rep. {\bf 254}, 215  (1995).

\bibitem{Meakin93}
P.~Meakin,
\newblock Phys. Rep. {\bf 235}, 189  (1993).

\bibitem{Family_Physica}
F.~Family and T.~Vicsek,
\newblock J. Phys. A {\bf 18}, L75 (1985).

\bibitem{KPZ86}
M.~Kardar, G.~Parisi, and Y.-C. Zhang,
\newblock Phys. Rev. Lett. {\bf 56}, 889 (1986).

\bibitem{VLDS_91}
Z.-W. Lai and S.~Das~Sarma,
\newblock Phys. Rev. Lett. {\bf 66}, 2348 (1991).

\bibitem{Meakin86}
P.~Meakin, P.~Ramanlal, L.~M. Sander, and R.~C. Ball,
\newblock Phys. Rev. A {\bf 34}, 5091 (1986).

\bibitem{plischke_prl_84}
M.~Plischke and Z.~R\'acz,
\newblock Phys. Rev. Lett. {\bf 53}, 415 (1984).

\bibitem{jullien_85}
R.~Jullien and R.~Botet,
\newblock Phys. Rev. Lett. {\bf 54}, 2055 (1985).

\bibitem{eden_plischke_85}
M.~Plischke and Z.~R\'acz,
\newblock Phys. Rev. A {\bf 32}, 3825 (1985).

\bibitem{Plischek_87}
M.~Plischke, Z.~R\'acz, and D.~Liu,
\newblock Phys. Rev. B {\bf 35}, 3485 (1987).

\bibitem{Vicsek_90}
T.~Vicsek, M.~Cserz{\H{o}}, and V.~K. Horv{\'a}th,
\newblock Physica A {\bf 167}, 315 (1990).

\bibitem{Family_Vicsek91}
F.~Family and T.~Vicsek (eds.),
\newblock {\em Dynamics of Fractal Surfaces},
\newblock (World Scientific, Singapore, New Jersey, 1991).

\bibitem{K_A_Takeuchi}
K.~A. Takeuchi and M.~Sano,
\newblock Phys. Rev. Lett. {\bf 104}, 230601 (2010).

\bibitem{kazumasa}
K.~A. Takeuchi, M.~Sano, T.~Sasamoto, and H.~Spohn,
\newblock Sci. Rep. {\bf 1}, 34 (2011).

\bibitem{saarloos_86}
W.~van Saarloos and G.~H. Gilmer,
\newblock Phys. Rev. B {\bf 33}, 4927 (1986).

\bibitem{goldenfeld}
N.~Goldenfeld,
\newblock J. Phys. A {\bf 17}, 2807 (1984).

\bibitem{krug_pra_88}
J.~Krug and H.~Spohn,
\newblock Phys. Rev. A {\bf 38}, 4271 (1988).

\bibitem{Michael_Herbert_00}
M.~Pr\"ahofer and H.~Spohn,
\newblock Phys. Rev. Lett. {\bf 84}, 4882 (2000).

\bibitem{forster_pra_16_732_77}
D.~Forster, D.~R. Nelson, and M.~J. Stephen,
\newblock Phys. Rev. A {\bf 16}, 732 (1977).

\bibitem{kuramoto2003chemical}
Y.~Kuramoto,
\newblock {\em Chemical Oscillations, Waves, and Turbulence},
\newblock (Dover Publications, 2003).

\bibitem{sivashinsky_Acta_77}
G.~I. Sivashinsky,
\newblock Acta. Astron. {\bf 4}, 1177 (1977).

\bibitem{fisher91}
D.~S. Fisher and D.~A. Huse,
\newblock Phys. Rev. B {\bf 43}, 10728 (1991).

\bibitem{kpz87}
M.~Kardar and Y.-C. Zhang,
\newblock Phys. Rev. Lett. {\bf 58}, 2087 (1987).

\bibitem{JMKIM}
J.~M. Kim, M.~A. Moore, and A.~J. Bray,
\newblock Phys. Rev. A {\bf 44}, 2345 (1991).

\bibitem{huse_85}
D.~A. Huse, C.~L. Henley, and D.~S. Fisher,
\newblock Phys. Rev. Lett. {\bf 55}, 2924 (1985).

\bibitem{huse85}
D.~A. Huse and C.~L. Henley,
\newblock Phys. Rev. Lett. {\bf 54}, 2708 (1985).

\bibitem{kerstein_92}
A.~R. Kerstein and W.~T. Ashurst,
\newblock Phys. Rev. Lett. {\bf 68}, 934 (1992).

\bibitem{KPZ89}
E.~Medina, T.~Hwa, M.~Kardar, and Y.-C. Zhang,
\newblock Phys. Rev. A {\bf 39}, 3053 (1989).

\bibitem{Baiod88}
R.~Baiod, D.~Kessler, P.~Ramanlal, L.~Sander, and R.~Savit,
\newblock Phys. Rev. A {\bf 38}, 3672 (1988).

\bibitem{Pfeifer_83}
P.~Pfeifer, D.~Avnir, and D.~Farin,
\newblock Surf. Sci. {\bf 126}, 569  (1983).

\bibitem{Moskovits_85}
M.~Moskovits,
\newblock Rev. Mod. Phys. {\bf 57}, 783 (1985).

\bibitem{Colaiori_PRL_86_3946}
F.~Colaiori and M.~A. Moore,
\newblock Phys. Rev. Lett. {\bf 86}, 3946 (2001).

\bibitem{Beijeren_PRL_54_2026}
H.~van Beijeren, R.~Kutner, and H.~Spohn,
\newblock Phys. Rev. Lett. {\bf 54}, 2026 (1985).

\bibitem{Hwa_PRA_44_R7873}
T.~Hwa and E.~Frey,
\newblock Phys. Rev. A {\bf 44}, R7873 (1991).

\bibitem{Lassig_PRL_80_2366}
M.~L\"assig,
\newblock Phys. Rev. Lett. {\bf 80}, 2366 (1998).

\bibitem{Schwartz_Edwards_1992}
M.~Schwartz and S.~F. Edwards,
\newblock Europhys. Lett. {\bf 20}, 301 (1992).

\bibitem{Canet_PRL_104_150601}
L.~Canet, H.~Chat\'e, B.~Delamotte, and N.~Wschebor,
\newblock Phys. Rev. Lett. {\bf 104}, 150601 (2010).

\bibitem{Kloss_PRE2012}
T.~Kloss, L.~Canet, and N.~Wschebor,
\newblock Phys. Rev. E {\bf 86}, 051124 (2012).

\bibitem{Huergo_radial}
M.~A.~C. Huergo, M.~A. Pasquale, P.~H. Gonz\'alez, A.~E. Bolz\'an, and A.~J.
  Arvia,
\newblock Phys. Rev. E {\bf 84}, 021917 (2011).

\bibitem{Huergo_10}
M.~A.~C. Huergo, M.~A. Pasquale, A.~E. Bolz\'an, A.~J. Arvia, and P.~H.
  Gonz\'alez,
\newblock Phys. Rev. E {\bf 82}, 031903 (2010).

\bibitem{Reis_pre_70_031603_05}
F.~D.~A. Aar\~ao Reis,
\newblock Phys. Rev. E {\bf 70}, 031607 (2004).

\bibitem{J-krug_92}
J.~Krug, P.~Meakin, and T.~Halpin-Healy,
\newblock Phys. Rev. A {\bf 45}, 638 (1992).

\bibitem{sasamoto_prl_104_230602_10}
T.~Sasamoto and H.~Spohn,
\newblock Phys. Rev. Lett. {\bf 104}, 230602 (2010).


\bibitem{2010JSMTE_11_013S}
T.~{Sasamoto} and H.~{Spohn},
\newblock J. Stat. Mech. P11013 (2010).

\bibitem{meanfield}
F.~Ginelli and H.~Hinrichsen,
\newblock J. Phys. A: Math. Gen. {\bf 37}, 11085 (2004).

\bibitem{yakhot_j_s_comput_1_3_86}
V.~Yakhot and S.~Orszag,
\newblock J. Sci. Comput. {\bf 1}, 3 (1986).

\bibitem{hohenberg_rmp_49_435_77}
P.~C. Hohenberg and B.~I. Halperin,
\newblock Rev. Mod. Phys. {\bf 49}, 435 (1977).

\bibitem{yakhot_prl_57_1772_86}
V.~Yakhot and S.~A. Orszag,
\newblock Phys. Rev. Lett. {\bf 57}, 1722 (1986).

\bibitem{frey_pre_50_1024_94}
E.~Frey and U.~C. T\"auber,
\newblock Phys. Rev. E {\bf 50}, 1024 (1994).

\bibitem{s.k.Ma_prb_11_4077_75}
S.-K. Ma and G.~F. Mazenko,
\newblock Phys. Rev. B {\bf 11}, 4077 (1975).

\bibitem{Deker}
U.~Deker and F.~Haake,
\newblock Phys. Rev. A {\bf 11}, 2043 (1975).

\bibitem{bartelt_jpa_26_2743_93}
M.~C. Bartelt and J.~W. Evans,
\newblock J. Phys. A. {\bf 26}, 2743  (1993).

\bibitem{calabrese_prl_106_250603_11}
P.~Calabrese and P.~Le~Doussal,
\newblock Phys. Rev. Lett. {\bf 106}, 250603 (2011).

\bibitem{jmkim_prl_62_2289_89}
J.~M. Kim and J.~M. Kosterlitz,
\newblock Phys. Rev. Lett. {\bf 62}, 2289 (1989).

\bibitem{Imamura_prl_108_2012}
T.~Imamura and T.~Sasamoto,
\newblock Phys. Rev. Lett. {\bf 108}, 190603 (2012).

\bibitem{Takeuchi_prl_110_2013}
K.~A. Takeuchi,
\newblock Phys. Rev. Lett. {\bf 110}, 210604 (2013).

\bibitem{Imamura_JSP_2013}
T.~Imamura and T.~Sasamoto,
\newblock J. Stat. Phys. {\bf 150}, 908 (2013).

\bibitem{Maunuksela_PRL_79_1515}
J.~Maunuksela, M.~Myllys, O.-P.~K\"ahk\"onen, J.~Timonen, 
N.~Provatas, M. J. Alava, and T.~Ala-Nissila,
\newblock Phys. Rev. Lett. {\bf 79}, 1515 (1997).

\bibitem{miettinena_epjb_46_55_05}
L.~Miettinena, M.~Myllys, J.~Merikoski, and J.~Timonen,
\newblock Eur. Phys. J. B {\bf 46}, 55 (2005).

\bibitem{coment_Takeuchi_12}
K.~A.Takeuchi,
\newblock preprint  (2012).

\end{thebibliography}
\end{document}